\documentclass[twocolumn,pra,aps]{revtex4}
\usepackage{psfrag}
\usepackage{epsfig}
\usepackage{amsmath}
\usepackage{amssymb}

\usepackage{dsfont}
\usepackage[colorlinks=true,linkcolor=blue,citecolor=blue]{hyperref}

\newcommand{\bra}[1]{\ensuremath{\langle#1|}}
\newcommand{\ket}[1]{\ensuremath{|#1\rangle}}

\newcommand{\be}{\begin{equation}}
\newcommand{\ee}{\end{equation}}

\newcommand{\tr}{\textrm{tr}}

\newcommand{\ie}{{\it i.e.}}

\newcommand{\eg}{{\it e.g.}}

\newcommand{\adop}{\hat a^{\dagger}}
\newcommand{\aop}{\hat a}
\newcommand{\bdop}{\hat b^{\dagger}}
\newcommand{\bop}{\hat b}

\begin{document}

\title{Optomechanics assisted with a qubit:\\ From dissipative state preparation to many-body physics}

\date{\today}

\author{Anika C. Pflanzer}
\email{anika.pflanzer@mpq.mpg.de}
\author{Oriol Romero-Isart }
\author{J. Ignacio Cirac}
\affiliation{Max-Planck-Institut f\"ur Quantenoptik,
Hans-Kopfermann-Strasse 1,
D-85748, Garching, Germany.}

\begin{abstract}

We propose and analyze nonlinear optomechanical protocols that can be implemented by adding a single atom to an optomechanical cavity. In particular, we show how to engineer the environment in order to dissipatively prepare the mechanical oscillator in a superposition of Fock states with fidelity close to one. Furthermore, we discuss how a single atom in a cavity with several mechanical oscillators can be exploited to realize nonlinear many-body physics by stroboscopically driving the mechanical oscillators. We show how to prepare non-classical many-body states by either applying coherent protocols or engineering dissipation. The analysis of the protocols is carried out using a perturbation theory for degenerate Liouvillians and numerical tools. Our results apply to other systems where a qubit is coupled to a mechanical oscillator via a bosonic mode, \eg, in cavity quantum electromechanics.

\end{abstract}

\maketitle

\section{Introduction}
Mechanical devices operating close to the quantum regime are becoming ubiquitous in fundamental and applied science~\cite{Marquardt09, Kippenberg08}. However, even after achieving the milestone of ground-state cooling~\cite{Teufel11, chan11}, many of the applications have yet to be implemented: using quantum nano- and micromechanical devices as universal quantum transducers~\cite{stannigel12, Rabl10, Hammerer09}, building ultra-high sensitivity detectors exploiting quantum metrology~\cite{cleland98, yang06, chiu08, Geraci10}, and performing tests of the foundations of quantum mechanics~\cite{marshall03, RomeroIsart11b,RomeroIsart11c, pikovski12, chen13}. Aside from circumventing dissipation, the main challenge remains the realization of nonlinearities required for the preparation of non-Gaussian states. Due to their negative Wigner functions, they are inherently different from classical states and are essential to the realization of most of the above-mentioned applications.

Despite the intrinsic nonlinearity of optomechanical interactions at the single-photon level, the resulting couplings are usually very small~\cite{chan11, verhagen12}. In most setups, the single-photon interaction is enhanced by strongly driving the light field~\cite{Marquardt09, Kippenberg08} at the price of rendering the coupling linear. Consequently, the resulting Hamiltonians are at most quadratic in the field operators and do not alter the character of an initially Gaussian state. Apart from developing methods to enhance single-photon couplings~\cite{nunnenkamp11,  stannigel12, rabl11}, a promising strategy to overcome this obstacle is to couple the mechanical oscillator to an auxiliary system that can be easily prepared in a non-Gaussian state. Possible candidates are, \eg, a single photon~\cite{romeroisart10, romeroisart2011a, akram10}, a superconducting qubit~\cite{Connell10}, or even an intrinsic two-level defect~\cite{ramos13}.  Along these lines, coupling the motion of a single atom to a membrane has been proposed~\cite{hammerer09b} and even experimentally demonstrated for a cloud of ultracold atoms~\cite{camerer11b}.

In this article we propose to add a single atom to the optomechanical cavity in order to couple the mechanical oscillator to its internal structure. This is motivated by the improved finesse of optomechanical cavities approaching the strong-coupling regime for single atoms~\cite{verhagen12, sankey10}.
We show that not only may coherent methods be applied to realize non-Gaussian physics, but that the strong decoherence through the cavity can prepare the nanomechanical oscillator in a non-Gaussian steady state with fidelity close to one. The main idea is to exploit the dissipation rather than treating it as an obstacle~\cite{poyatos96}. While this approach has been proposed to prepare squeezed and entangled states of mechanical oscillators~\cite{mancini03, rabl04, vitali07, genes08, hartmann08, ludwig10, muschik11, wang13, tan13}, here we show how to use it to prepare non-Gaussian states. We further discuss how many-body physics can be implemented by adding $N$ mechanical oscillators into a cavity containing a single atom. A system with $N$ nonlinear modes is realized by stroboscopically driving the oscillators' frequencies. Using the time-dependence of the mechanical frequencies, we show how to prepare many-body non-Gaussian states using both dissipative and coherent methods.
The results presented here are applicable to the general case where a single qubit is coupled to a mechanical oscillator via a bosonic mode. This can be achieved in a variety of physical systems, \eg, in cavity quantum electromechanics~\cite{Teufel11, Connell10}. 

The manuscript is organized as follows: in Sec.~\ref{sec:setup} we describe the system, list the assumptions, and define the Hamiltonian. In Sec.~\ref{sec:diss} we present the main result of the manuscript: the dissipative preparation of the system in a non-Gaussian state. In Sec.~\ref{sec:pert} we describe a general perturbation theory for degenerate Liouvillians that is used to explain the numerical results. In Sec.~\ref{sec:noise} we show how the fidelity to prepare a non-Gaussian state depends on the system's inherent noise. In Sec.~\ref{sec:jump} the fidelity is optimized by engineering the environment. Insights from perturbation theory are given in Sec.~\ref{sec:explain}. In Sec.~\ref{sec:coh} coherent methods for state preparation are described. The extension to many-body systems is given in Sec.~\ref{sec:many}, where both dissipative (Sec.~\ref{sec:many_prep}) and coherent approaches (Sec.~\ref{sec:law_many}) are analyzed. Finally, we draw the conclusions and give an outlook in Sec.~\ref{sec:conc}.

\section{The setup}\label{sec:setup} 
We consider a two-level system and a mechanical oscillator both coupled to a cavity. The system's Hamiltonian is given by (we set $\hbar=1$ throughout the manuscript)
\be
\begin{split}
H=& \Delta \adop_1\aop_1+\frac{\delta}{2}\hat{\sigma}_z+\omega\bdop\bop+g_{\rm m}(\adop_1\bop+\aop_1\bdop)\\
&-g_{\rm q}(\aop_1\hat{\sigma}^++\adop_1\hat{\sigma}^-)+\Omega(\hat{\sigma}^++\hat{\sigma}^-)+H_{\rm aux},
\label{eq:ham}
\end{split}
\ee 
with 
\be
H_{\rm aux}=\Delta^{\rm aux}\adop_2\aop_2-g_{\rm m}^{\rm aux}(\adop_2\bdop+\aop_2\bop)+g_{\rm q}^{\rm aux}(\aop_2^{\dagger}\hat{\sigma}^++\aop_2\hat{\sigma}^-).
\ee
$\bop (\bdop)$ describe the annihilation (creation) operators of the mechanical mode at frequency $\omega$. We assume that the cavity supports two modes with annihilation (creation) operators $\aop_i (\adop_i)$ $(i=1,2)$ detuned by $\Delta$ and $\Delta^{\rm aux}$ respectively. Both modes are strongly driven, $\aop_1$ ($\aop_2$) with a red (blue)-detuned field, such that their single-photon coupling strength is enhanced by the square root of the number of steady-state photons to $g_{\rm m}$ ($g_{\rm m}^{\rm aux}$). The qubit is described by the lowering (raising) operators $\hat{\sigma}^- (\hat{\sigma}^+)$ detuned from the laser frequency by $\delta$, strongly driven at $\Omega$, and coupled to the two cavity modes by $g_{\rm q}$ and $g_{\rm q}^{\rm aux}$ respectively. 

The dissipative processes are described by master equations of Lindblad form. The loss of cavity photons with a decay rate $\kappa$ is given by
\be
\mathcal{L}_{ \rm cav}[\rho]=2\kappa\left[\aop_1\rho \adop_1-\frac{1}{2}\{\adop_1 \aop_1, \rho\}_+\right].
\ee
The decay of the auxiliary mode $\aop_2$ is defined in full analogy with decay rate $\kappa_{\rm aux}$. The dissipation caused by the qubit is given by 
\be
\mathcal{L}_{\rm q}[\rho]=\gamma_{\rm q}\left[\hat{\sigma}^- \rho \hat{\sigma}^+-\frac{1}{2}\{\hat{\sigma}^+ \hat{\sigma}^-, \rho\}_+\right],\label{eq:lq}
\ee
where $\gamma_{\rm q}$ is the spontaneous emission rate. For the mechanical oscillator the decoherence at a rate $\gamma_{\rm m}$ is described by 
\be
\mathcal{L}_{\rm m}[\rho]=\gamma_{\rm m}\left[(\bop+\bdop) \rho (\bdop+\bop)-\frac{1}{2}\{(\bop+\bdop)^2, \rho\}_+\right].\label{eq:lm}
\ee
We choose decoherence of the localization type~\cite{romeroisart2011a, pflanzer12}, \eg, dominant in levitating dielectrics. For a different decoherence mechanism, the analysis is in full analogy.

Throughout the article we consider the regime where the cavity merely mediates the interaction between the oscillator and the two-level system, and can be adiabatically eliminated. Therefore, the following conditions have to be fulfilled: first, the coupling between the cavity and both the oscillator and the qubit has to be small, fulfilling either $g_{\rm q(m)}/\kappa\ll1$ (dissipative dynamics, see Sec.~\ref{sec:diss}), or $g_{\rm q}/|\Delta-\delta|\ll1,g_{\rm m}/|\Delta-\omega|\ll1 $ (coherent dynamics, see Sec.~\ref{sec:coh}), or  both conditions. Second, the interaction mediated by the cavity has to be stronger than the dissipative processes leading to the good-cooperativity requirement for both the qubit $\mathcal{C}_{\rm q}=g_{\rm q}^2/( \kappa \gamma_{\rm q})>1$, and the mechanical oscillator $\mathcal{C}_{\rm m}=g_{\rm m}^2/( \kappa \gamma_{\rm m})>1$. Note that the more demanding strong-coupling limit, $g_{\rm m}>\gamma_{\rm m}, \kappa$ and $g_{\rm q}>\gamma_{\rm q}, \kappa$ is not necessary (the same conditions apply to the cavity mode $\aop_2$).

Possible realizations of Hamiltonian eq.~\eqref{eq:ham} range from electromechanical setups~\cite{Teufel11, teufel11a}, where a microresonator couples a mechanical oscillator to a superconducting qubit, to cavity-optomechanical systems with a cavity mediating the interaction between a two-level atom and a mechanical membrane~\cite{sankey10, hammerer09b, camerer11b} or a levitating sphere~\cite{romeroisart10, chang10, romeroisart2011a}. 
Remarkably, in the specific case of levitating spheres, the regime where ground-state cooling is possible makes the same cavity suitable for coupling to single atoms~\cite{romeroisart2011a}. This is due to the fact that in this case, the cooperativity of the mechanical oscillator reduces to the single-atom case $\mathcal{C}_{\rm m}=\mathcal{C}_{\rm q}$ and only depends on cavity parameters~\footnote{Light-induced dissipation processes dominate $\gamma_{\rm m}$ reducing the cooperativity of the nanomechanical resonator to the one of the single-atom case entirely determined by the cavity, $\mathcal{C}_{\rm m}=\mathcal{C}_{\rm q}=c^3/(2 \omega_{\rm c}^2 V_{\rm c}\kappa)$, where $V_{\rm c}$ is the cavity volume and $c$ the speed of light. The minimal phonon number attainable when cooling a mechanical oscillator in the resolved sideband regime is given by $n_{\rm min}=(\kappa/(4\omega))^2+1/(4 \mathcal{C}_{\rm m})$, thus the two conditions are equivalent.}.

\section{Dissipative dynamics}\label{sec:diss}
The goal of preparing non-Gaussian states of nano-mechanical oscillators is often hindered by the unavoidable occurrence of dissipation. In contrast, the proposed protocol exploits the interaction with the environment to prepare a mechanical oscillator in a non-Gaussian dark state with fidelity close to one. This goes along the line of ideas developed and analyzed recently for a variety of different systems~\cite{poyatos96, muschik11, wang13, stannigel12a}. 
We assume the limit where dissipation dominates, namely $g_{\rm q(m)}/\kappa\ll1$, and choose  $g_{\rm m}=g_{\rm q}$, $\Omega=0$, $\Delta=\delta=\omega$, and $H_{\rm aux}=0$. An adiabatic elimination of the cavity mode in the Hamiltonian, eq.~\eqref{eq:ham}, yields an effective dissipative dynamics governed by the Liouvillian
\be
\begin{split}
\mathcal{L}_{0}[\rho]&=\gamma_{\rm eff}\left[\hat{J}_0\rho J_0^{\dagger}-\frac{1}{2}\{\hat{J}_0^{\dagger}\hat{J}_0, \rho\}_+\right].
\end{split}\label{eq:leff}
\ee
Here, the jump operator is given by $\hat{J}_0=\bop-\hat{\sigma}^-$ and the effective decay rate by $\gamma_{\rm eff}=2 g_{\rm m}^2/\kappa$.  
$\mathcal{L}_0[\rho]$ possesses two degenerate steady states, 
\be
\rho_{\rm A}=\frac{1}{2}\left(|g,1\rangle+|e,0\rangle\right)\left(\langle g,1|+\langle e,0|\right),
\ee
and $\rho_{\rm B}=|g,0\rangle\langle g,0|$.
Here, $g$ $(e)$ describes the qubit's ground (excited) state in the basis of $\hat{\sigma}_z$, and $0$ $(1)$ the ground (excited) state of the phononic mode. While $\rho_{\rm A}$ is a non-Gaussian entangled state for the phonon, $\rho_{\rm B}$ describes the Gaussian ground state. This degeneracy can be lifted by additional dissipative terms and is very sensitive to any perturbation, as shown below. The goal is to lift the degeneracy such that the probability to prepare $\rho_{\rm A}$ is maximized. To achieve this, we introduce a perturbation theory for degenerate Liouvillians in Sec.~\ref{sec:pert}. Following this, we investigate the steady states including the noise operators $\mathcal{L}_{\rm q}$ (eq.~\eqref{eq:lq}) and $\mathcal{L}_{\rm m}$ (eq.~\eqref{eq:lm}) in Sec.~\ref{sec:noise}. In Sec.~\ref{sec:jump}, an additional general linear jump operator is introduced and specified such that the probability to prepare the non-Gaussian state is maximized. In Sec.~\ref{sec:explain}, the analysis is completed by a consideration of the perturbative regime that explains the results.
\subsection{Perturbation theory for degenerate Liouvillians}\label{sec:pert}
In the following, we give a description of the perturbation theory for degenerate Liouvillians~\cite{poyatos96} used throughout the paper. In order to determine the steady state of a Liouvillan described by $\mathcal{L}=\mathcal{L}_0+\epsilon\mathcal{L}_{\rm pert}$, with $\epsilon\ll1$, we can treat $\mathcal{L}_{\rm pert}$ as a perturbation to $\mathcal{L}_0$. The underlying concept is to provide an effective description of the dynamics of the fast subspace (given by $\mathcal{L}_{\rm pert}$) by applying a transformation that dresses the eigenstates of the slow subspace (given by $\mathcal{L}_0$).
An expansion of the effective Liouvillian in terms of the perturbation parameter $\epsilon$ yields
\be
\mathcal{L}_{\rm eff}=\mathcal{L}_0+\epsilon\mathds{P}\mathcal{L}_{\rm pert}\mathds{P}-\epsilon^2\mathds{P}\mathcal{L}_{\rm pert}\mathds{Q}\mathcal{L}_0^{-1}\mathds{Q}\mathcal{L}_{\rm pert}\mathds{P}+...., \label{eq:per_suppl}
\ee
where $\mathds{P}$ ($\mathds{Q}=\mathds{1}-\mathds{P}$) projects into the subspace that is kept (eliminated). In the following, we show how to determine $\mathds{P}$. We define
\be
\mathds{P}=\rho_{\rm A}\otimes \chi_{\rm A}+\rho_{\rm B}\otimes\chi_{\rm B}.
\ee
Its action on an arbitrary density matrix $\mu$ is given by
\be
\mathds{P}\mu=\rho_{\rm A}\rm{tr}(\chi_{\rm A}\mu)+\rho_{\rm B}\rm{tr}(\chi_{\rm B}\mu).
\ee
Here, $\rho_i$ ($\chi_i$) (with $i=\rm{A,B}$) denote right (left) eigenvectors of the Liouvillian $\mathcal{L}_0$ ($\mathcal{L}_0^{\diamondsuit}$) with eigenvalue zero, where $\mathcal{L}_0^{\diamondsuit}$ denotes the Liouville operator acting on left states. That is, $\mathcal{L}_0[\rho_{\rm A (B)}]=0$, ($\chi_{\rm A (B)}\mathcal{L}_0^{\diamondsuit}=0$). Besides, for $\mathds{P}$ to be a projector, $\mathds{P}\left(\mathds{P}\rho\right)=\mathds{P}\left(\rho\right)$ and the completeness relation $\sum_{i,j=A, B}\rho_i\otimes\chi_j=1$ have to be fulfilled. This imposes biorthonormality, $\rm{tr}\left(\chi_{\rm A}\rho_{\rm B}\right)=\rm{tr}\left(\chi_{\rm B}\rho_{\rm A}\right)=0$ and $\rm{tr}\left(\chi_{\rm A}\rho_{\rm A}\right)=\rm{tr}\left(\chi_{\rm B}\rho_{\rm B}\right)=1$. Since the definition of $\mathds{P}$ is not unique due to the degeneracy of the Liouvillian $\mathcal{L}_0$, we impose the additional condition on 
\be
\mathds{P}\mathcal{L}_{\rm pert}\mathds{P}=\sum_{i,j=\rm{A,B}}\rm{tr}\left(\chi_{\rm i}\mathcal{L}_{\rm pert}[\rho_{\rm j}]\right)\rho_{\rm i}\otimes \chi_{j}
\ee
to be diagonal, \ie, $\rm{tr}\left(\chi_{\rm i}\mathcal{L}_{\rm pert}[\rho_{\rm j}]\right)=0$ for $i\neq j$. This is analogous to degenerate perturbation theory in the Hamiltonian case. 

The steady state of the Liouvillian in perturbation theory to first order is thus given by the eigenstate of $\mathcal{L}_0+\epsilon\mathds{P}\mathcal{L}_{\rm pert}\mathds{P}$ with eigenvalue zero. It can be shown that $\mathcal{L}_0+\epsilon\mathds{P}\mathcal{L}_{\rm pert}\mathds{P}$ with Liouvillians of Lindblad form always possesses a zero eigenvalue. To prove this, it is sufficient to demonstrate $\rm{tr}\left[(\mathcal{L}_0+\epsilon\mathds{P}\mathcal{L}_{\rm pert}\mathds{P})[\mu]\right]=0$~\footnote{Given the spectrum of $\mathcal{L}_{\rm eff}$ with the real part of all eigenvalues smaller or equal to zero, and the preservation of the trace, its direct consequence is the existence of a steady state.}. The Lindblad form of $\mathcal{L}_0$ and the trace's invariance under cyclic permutations yields $\rm{tr}\left(\mathcal{L}_0[\mu]\right)=0$. Furthermore,
\be
\begin{split}
\rm{tr}\left(\mathds{P}\mathcal{L}_{\rm pert}\mathds{P}[\mu]\right)&=\sum_{i,j=\rm{A,B}}\rm{tr}\left(\chi_{\rm i}\mathcal{L}_{\rm pert}[\rho_{\rm j}]\right)\rm{tr}\left(\chi_{j}\mu\right)\\
&=\sum_{j=\rm{A,B}}\rm{tr}\left(\chi_j\mu\right)\rm{tr}\Big(\mathcal{L}_{\rm pert}[\rho_j]\underbrace{\sum_{i=\rm{A,B}}\chi_i}_{=\mathds{1}}\Big)\\
&=0,
\end{split}
\ee
where the completeness of the left eigenvectors $\sum_i\chi_i=1$ and the Lindblad form of $\mathcal{L}_{\rm pert}$ have been used.

Higher orders of the perturbation theory can be determined analogously, but we will restrict the analysis to the lowest order in $\epsilon$ throughout this article. 
\subsection{Steady state with noise}\label{sec:noise}
\begin{figure}
\includegraphics[width=1.0\linewidth]{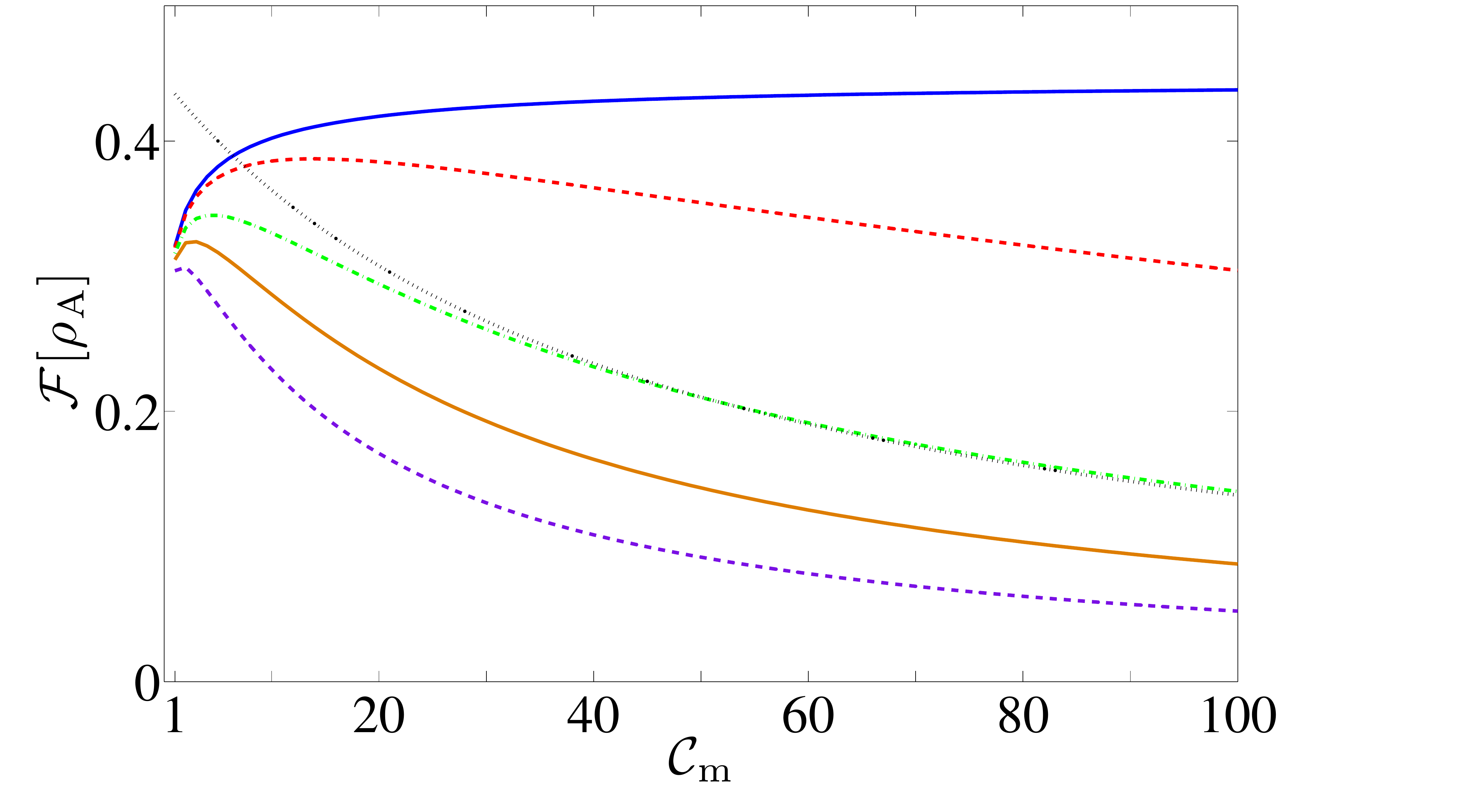}
\caption{(Color online) Fidelity for the preparation of the non-Gaussian state $\rho_{\rm A}$ as a function of $\mathcal{C}_{\rm m}$ for different qubit cooperativities $\mathcal{C}_{\rm q}$. \textit{Solid blue}: $\mathcal{C}_{\rm q}=\infty$, \textit{Dashed red}: $\mathcal{C}_{\rm q}=100$, \textit{Dash-dotted green:} $\mathcal{C}_{\rm q}=20$ (\textit{Dotted black:} comparison to the analytic result for $\mathcal{C}_{\rm q}=20$), \textit{Solid orange:} $\mathcal{C}_{\rm q}=10$, \textit{Dashed purple:} $\mathcal{C}_{\rm q}=5$.}\label{Fig1}
\end{figure}
We analyze the effect of the additional noise caused by the spontaneous decay of the atom $\mathcal{L}_{\rm q}$ (eq.~\eqref{eq:lq}) and the decoherence of the mechanical oscillator $\mathcal{L}_{\rm m}$ (eq.~\eqref{eq:lm}). These additional Liouvillians lift the original degeneracy of the steady state of $\mathcal{L}_0$. Perturbation theory to first order yields the unique dark state
\be
\rho_{\rm SS}=\alpha_{\rm n} \rho_{\rm A}+\beta_{\rm n} \rho_{\rm B}
\ee
for $\mathcal{L}_0+\mathds{P}\left(\mathcal{L}_{\rm m}+\mathcal{L}_{\rm q}\right)\mathds{P}$.
The coefficients depend on the noise parameters and are given by $\alpha_{\rm n}=4\gamma_{\rm m}/(4\gamma_{\rm q}+9\gamma_{\rm m})$ and $\beta_n=(4\gamma_{\rm q}+5\gamma_{\rm m})/(4\gamma_{\rm q}+9\gamma_{\rm m})$. To complement the analytical study, we carry out a numerical evaluation of the steady state, which is shown to be in good agreement with the perturbation theory for $\mathcal{C}_{\rm q}, \mathcal{C}_{\rm m}\gg1$, as illustrated in Fig.~\ref{Fig1}. As expected from the analytical result, the fidelity to prepare the entangled non-Gaussian state $\rho_{\rm A}$ is maximized for $\gamma_{\rm q}=0$ and can reach $\mathcal{F}[\rho_{\rm A}]=\tr[\rho_\text{SS} \rho_\text{A}] =\alpha_{\rm n}=4/9$. The optimal value of $\mathcal{C}_{\rm m}$ to maximize $\mathcal{F}[\rho_{\rm A}]$ for a given $\mathcal{C}_{\rm q}$ can be read from Fig.~\ref{Fig1}. Thus, the system's inherent noise leads to the preparation of a mechanical oscillator in a non-Gaussian state with a fidelity $\mathcal{F}[\rho_{\rm A}]\leq 4/9$.

\subsection{Steady state with an engineered environment}\label{sec:jump}
\begin{figure}
\includegraphics[width=1.0\linewidth]{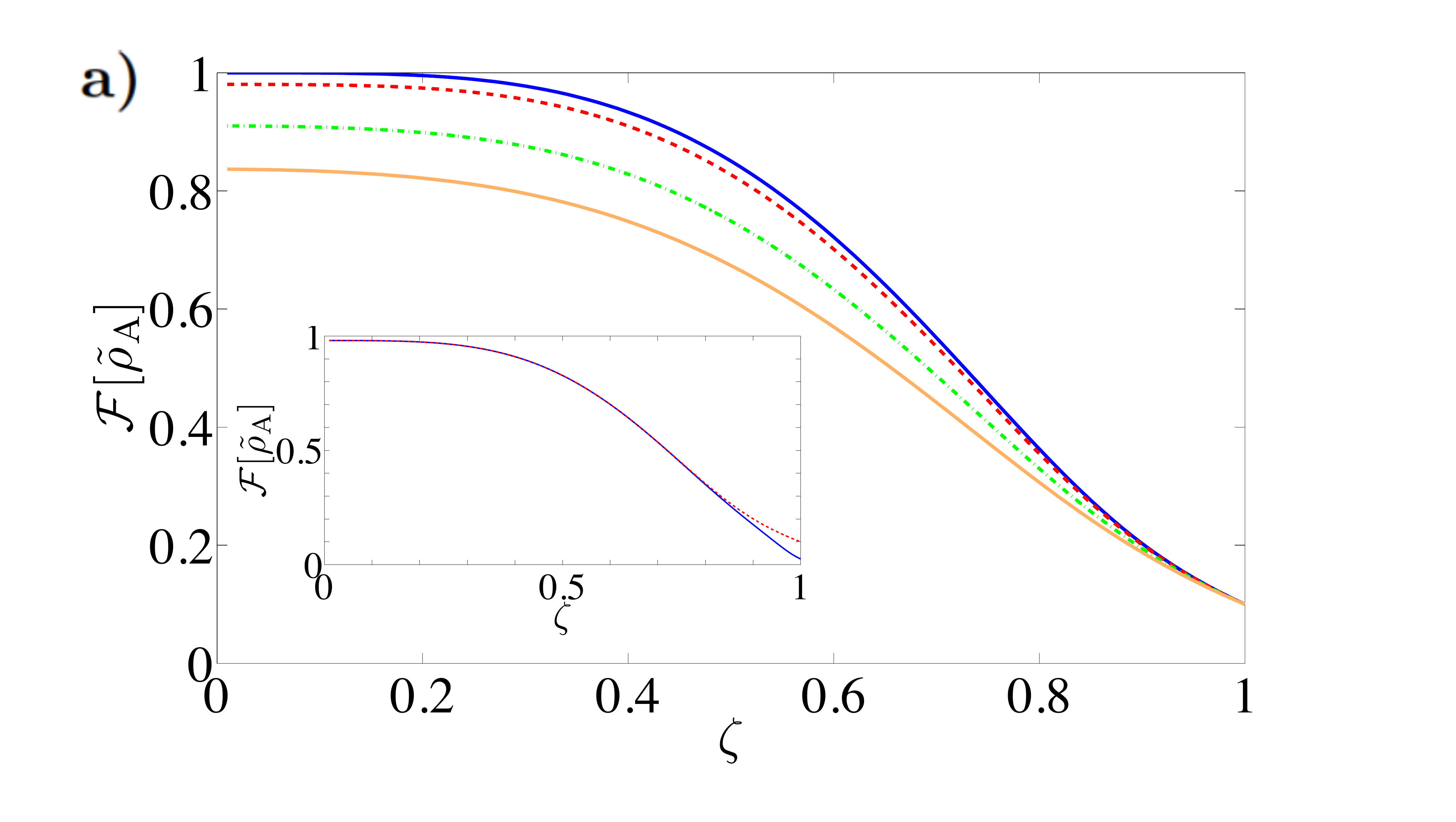}
\includegraphics[width=1.0\linewidth]{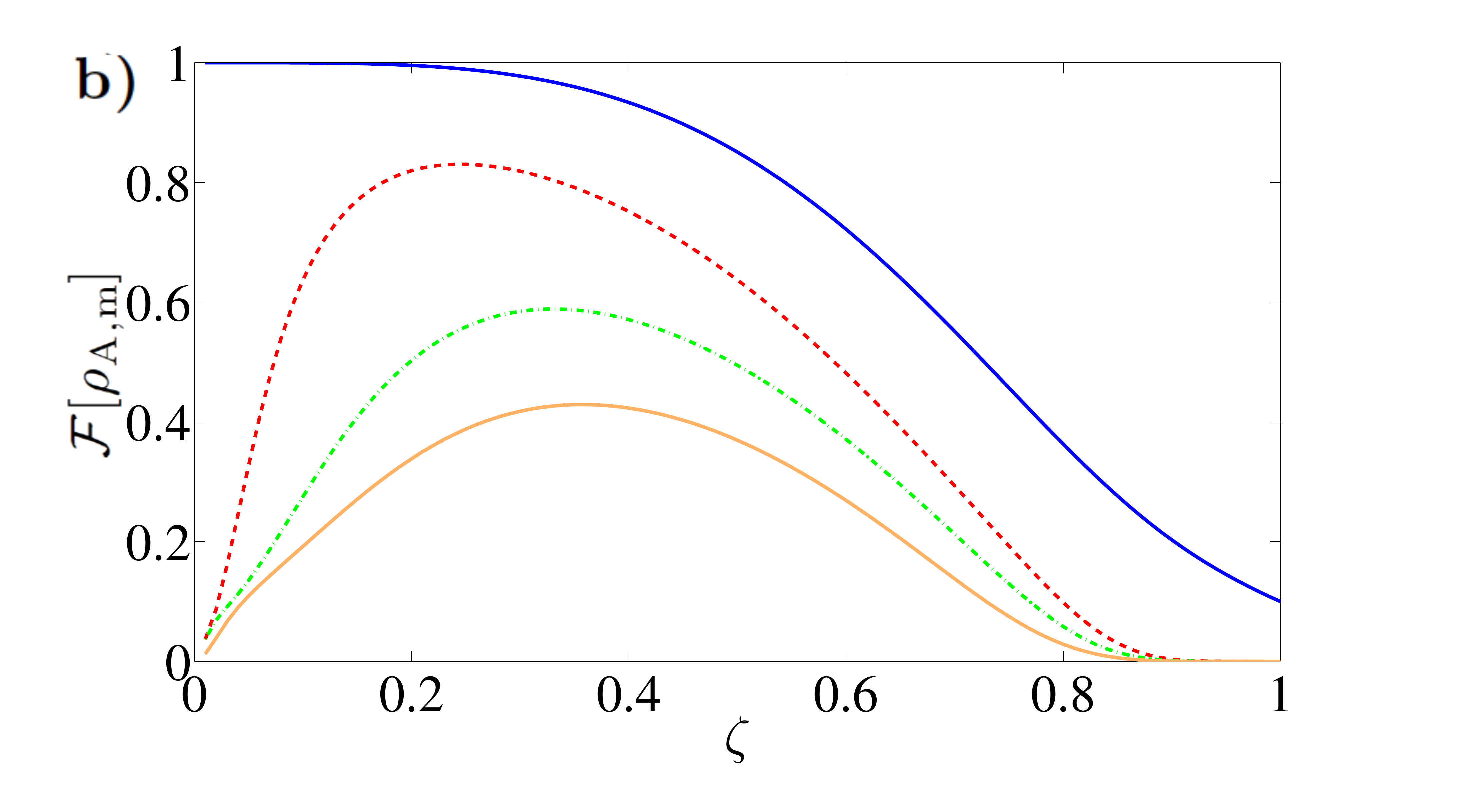}
\includegraphics[width=1.0\linewidth]{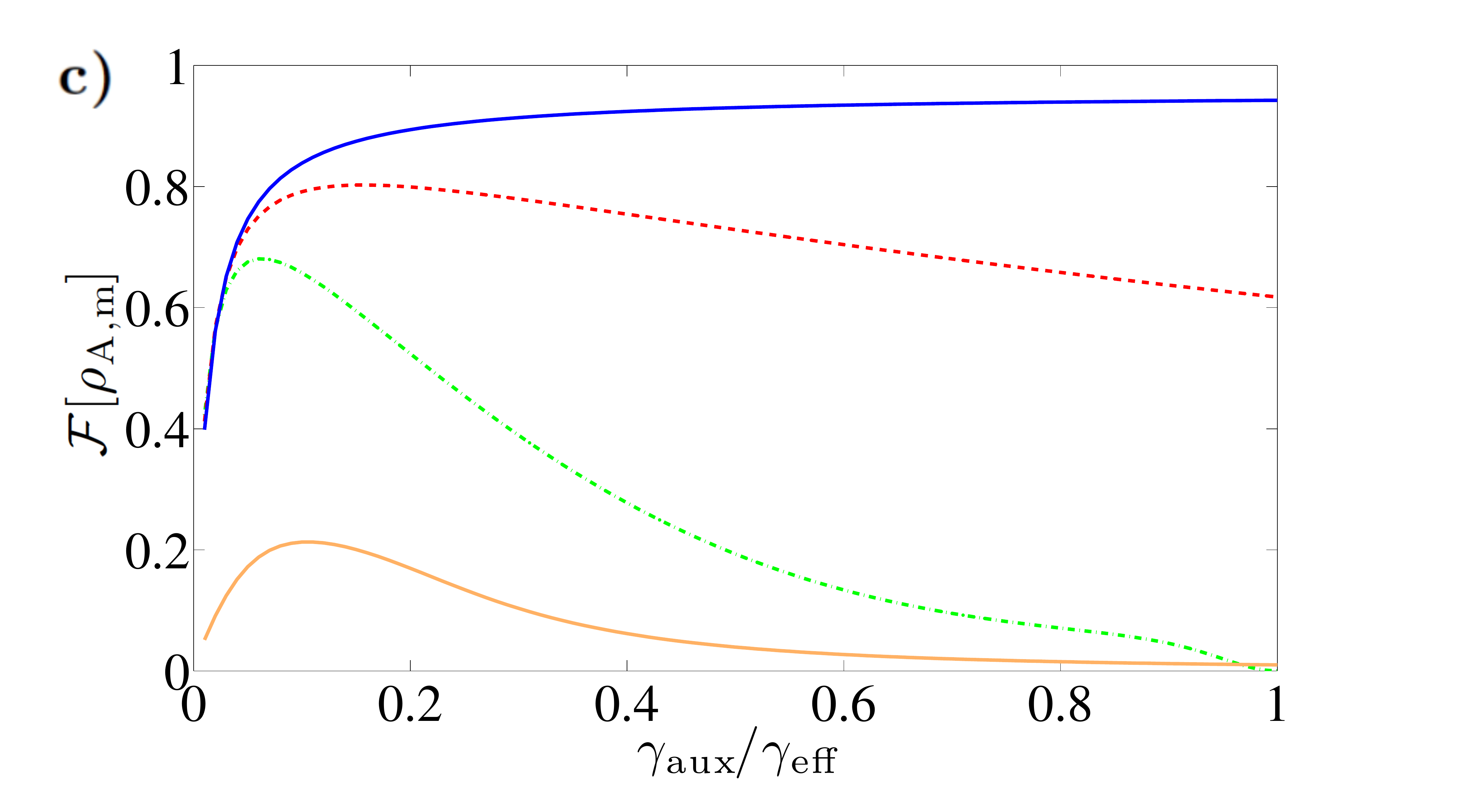}
\caption{(Color online) 
 Fidelity to prepare a) $\tilde{\rho}_{\rm A}$ and b) $\rho_{\rm A,m}$ as functions of $\zeta$ for $\gamma_{\rm aux}/\gamma_{\rm eff}=1$ and different cooperativities. The size of the Hilbert space for the mechanical oscillator is chosen as $N=10$. \textit{Solid blue}: $\mathcal{C}_{\rm q}=\mathcal{C}_{\rm m}=\infty$, \textit{Dashed red}: $\mathcal{C}_{\rm q}=\mathcal{C}_{\rm m}=100$,  \textit{Dash-dotted green}: $\mathcal{C}_{\rm q}=\mathcal{C}_{\rm m}=20$,  \textit{Solid orange}: $\mathcal{C}_{\rm q}=\mathcal{C}_{\rm m}=10$. \textit{Inset:} Comparison of the simulation for $\mathcal{C}_{\rm m}=\mathcal{C}_{\rm q}=100$ and different sizes of the Hilbert space. \textit{Solid blue:} $N=30$, \textit{Dashed red:} $N=10$. c) Fidelity to prepare $\rho_{\rm A, m}$ as a function of $\gamma_{\rm aux}/\gamma_{\rm eff}$ for $\zeta=0.2$, $\mathcal{C}_{\rm m}=\mathcal{C}_{\rm q}=1000$ and different jump operators $\hat{J}_1$. \textit{Solid blue:} $\hat{J}_1=\hat{\sigma}^+-\zeta \bdop$, \textit{Dashed red:} $\hat{J}_1=\hat{\sigma}^+$, \textit{Dash-dotted green:} $\hat{J}_1=\hat{\sigma}^++\zeta\bdop$, \textit{Solid orange:} $\hat{J}_1=\bdop$. }\label{Fig3}
\end{figure}
In the following we propose a protocol to enhance the fidelity for the preparation of non-Gaussian states. For this purpose, we consider the modified jump operator $\tilde{J}_0=\bop-\zeta \hat{\sigma}^-$ for $\mathcal{L}_0$ (given by eq.~\eqref{eq:leff}) with $\zeta= (g_{\rm q}/g_{\rm m})^2$ (we choose $\zeta\leq1$). It can be realized with the Hamiltonian eq.~\eqref{eq:ham} for $g_{\rm m}\neq g_{\rm q}$. The steady state of $\mathcal{L}_0$ is thus degenerate and composed of $\rho_{\rm B}$ as defined previously and
\be
\tilde{\rho}_{\rm A}=\frac{1}{1+\zeta^2}(\zeta |g,1\rangle+|e,0\rangle)(\zeta\langle g,1|+\langle e,0|).
\ee
In order to lift the degeneracy in a way that leads to an increased population in $\tilde{\rho}_{\rm A}$, we introduce an additional Liouvillian
\be
\mathcal{L}_{\rm aux}=\gamma_{\rm aux}\left[\hat{J}_1\rho\hat{J}_1^{\dagger}-\frac{1}{2}\{\hat{J}_1^{\dagger}\hat{J}_1, \rho\}_+\right],\label{eq:laux}
\ee
with jump operator $\hat{J}_1= \hat{\sigma}^+-\zeta\bdop$. This jump operator can be realized by including $H_{\rm aux}\neq 0$ in the Hamiltonian of eq.~\eqref{eq:ham} with a blue detuning $\Delta^{\rm aux}=-\omega=-\delta$. This yields $\gamma_{\rm aux}=2~\left(g_{\rm q}^{\rm aux}\right)^2/\kappa_{\rm aux}$ and $\zeta=(g_{\rm m}^{\rm aux}/g_{\rm q}^{\rm aux})^2$. Together with the noise terms $\mathcal{L}_{\rm m}$ and $\mathcal{L}_{\rm q}$, the steady state is given by
\be
\tilde{\rho}_{\rm SS}=\alpha_{\rm aux}\tilde{\rho}_{\rm A}+\beta_{\rm aux}\rho_{\rm B}.\label{eq:rhossaux}
\ee

In the presence of the inherent noise, the fidelity to prepare the system in the entangled non-Gaussian state $\tilde \rho_\text{A}$ is strongly enhanced by $\mathcal{L}_{\rm aux}$ as shown in Fig.~\ref{Fig3} a)~\footnote{Some care has to be taken in the numerical study as the system only exhibits a steady state for $\zeta <1$: in the regime where the bosonic operator $\bdop$ dominates, no steady state is reached. Consequently, the Hilbert space for the phononic mode needs to be sufficiently large, as a finite Hilbert space generally might yield a steady state although it does not exist. This is illustrated in the inset of Fig.~\ref{Fig3}~a), where the steady state obtained for a Hilbert space of size $N=10$  is compared to $N=30$. It demonstrates that in the regime of interest, namely where the fidelity to prepare $\tilde{\rho}_{\rm A}$ is high, they are in good agreement and the numerical study is valid.}. For example, for $\mathcal{C}_{\rm m}=\mathcal{C}_{\rm q}=100$ and $\zeta=0.2$, the fidelity for the preparation of $\tilde \rho_\text{A}$ is close to one, $\mathcal{F}[\tilde{\rho}_{\rm A}]=0.98$. Even for much smaller cooperativies, \eg, for $\mathcal{C}_{\rm m}=\mathcal{C}_{\rm q}=10$, the fidelity is $\mathcal{F}[\tilde{\rho}_{\rm A}]=0.82$. 

Despite the increment of the fidelity for the preparation of $\tilde \rho_\text{A}$, the amount of entanglement of the steady state depends on $\zeta$. For small $\zeta$, the state is close to the ground state of the harmonic oscillator and shows only little entanglement. To prevent this, we propose to measure the qubit in the basis $|+\rangle_{\rm q}=\left(\zeta|e\rangle+|g\rangle\right)/\sqrt{1+\zeta^2}, |-\rangle_{\rm q}=\left(|e\rangle-\zeta|g\rangle\right)/\sqrt{1+\zeta^2}$, and postselect to keep only the $|+\rangle_{\rm q}$-result. This prepares the mechanical oscillator in
\be
\rho_{\rm SS, m}=
\alpha_{\rm m}\rho_{\rm A,m}+\beta_{\rm m}\rho_{\rm B,m},\label{eq:fock}
\ee
with $\rho_{\rm A, m}=(|0\rangle+|1\rangle)(\langle 0|+\langle 1|)/2$ and $\rho_{\rm B, m}=|0\rangle\langle0|$.
In Fig.~\ref{Fig3}~b), we show that the maximal fidelity is $\mathcal{F}[\rho_{\rm A,m}]=\alpha_{\rm m}=0.83$ for $\zeta=0.25$ and cooperativities $\mathcal{C}_{\rm m}=\mathcal{C}_{\rm q}=100$. In comparison, when only the system's inherent noise is included, the maximal fidelity is $\mathcal{F}[\rho_{\rm A}]=4/9$ for $\mathcal{C}_{\rm m}=\mathcal{C}_{\rm q}=\infty$. In full analogy, Fock states can be prepared via a suitable choice of the measurement basis. For instance, by measuring in the $|g\rangle$ and $|e\rangle$-basis and postselecting to keep the $|g\rangle$-result, we can prepare the $|1\rangle$-state for the mechanical oscillator. For $\zeta=0.25$ and cooperativities $\mathcal{C}_{\rm q}=\mathcal{C}_{\rm m}=100$, a fidelity of $\mathcal{F}\approx 0.83$ is achievable . 

Furthermore, we investigate the dependence of $\mathcal{F}[\rho_{\rm A, m}]$ on $\gamma_{\rm aux}/\gamma_{\rm eff}$ as shown in Fig.~\ref{Fig3}c). We also analyze different jump operators $\hat{J}_1$ and demonstrate that the optimal configuration to maximize $\mathcal{F}[\rho_{\rm A, m}]$ is achieved for $\hat{J}_1=\hat{\sigma}^+-\zeta \bdop$ and $\gamma_{\rm aux}/\gamma_{\rm eff}\approx 1$. Note that throughout this subsection we rely on numerical simulations since the perturbation theory of Sec.~\ref{sec:pert} is only valid in the regime $\gamma_{\rm aux}/\gamma_{\rm eff}\ll1$. 
\subsection{Insights from perturbation theory}\label{sec:explain}
\begin{figure}
\includegraphics[width=1.0\linewidth]{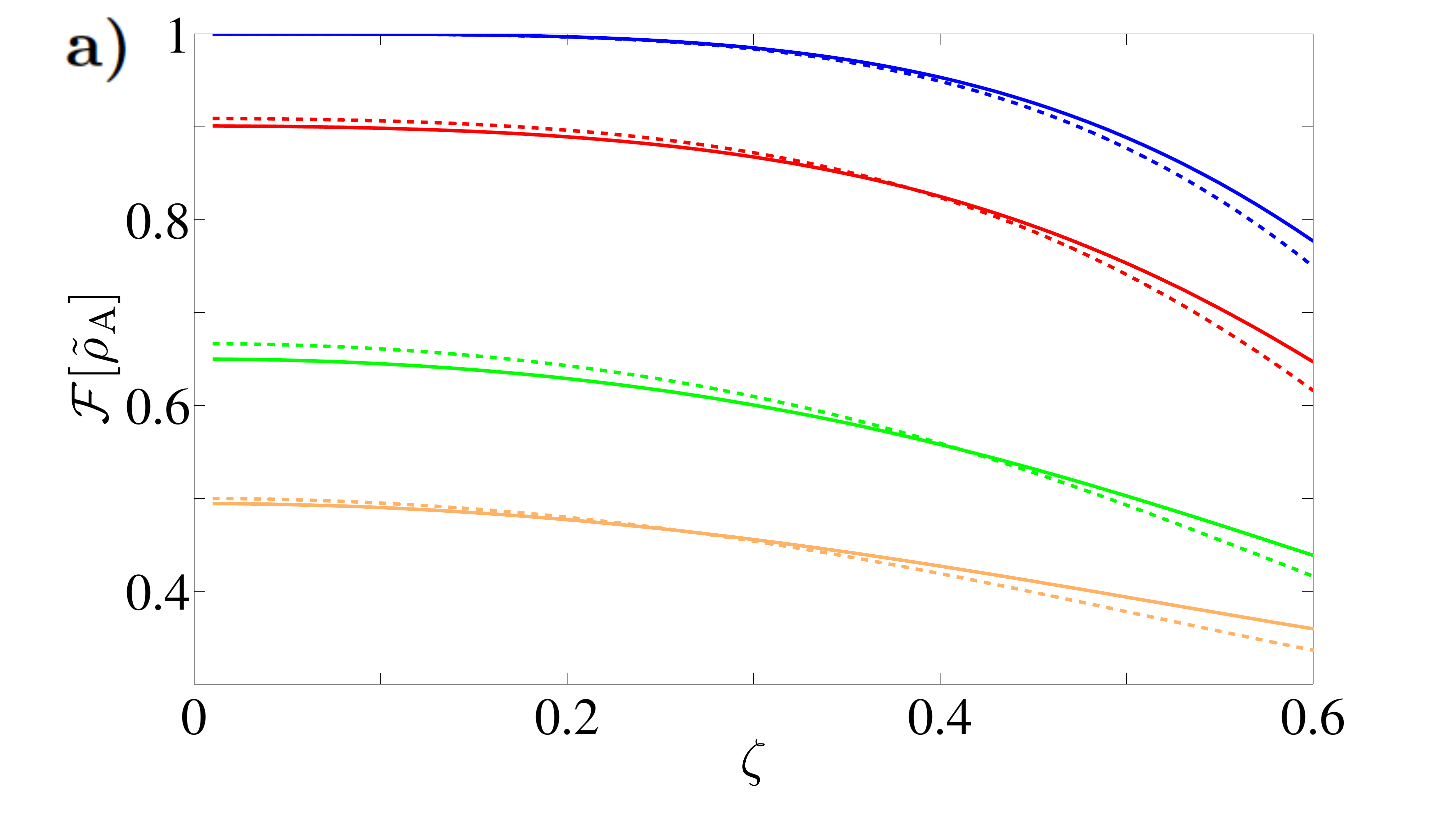}
\includegraphics[width=1.0\linewidth]{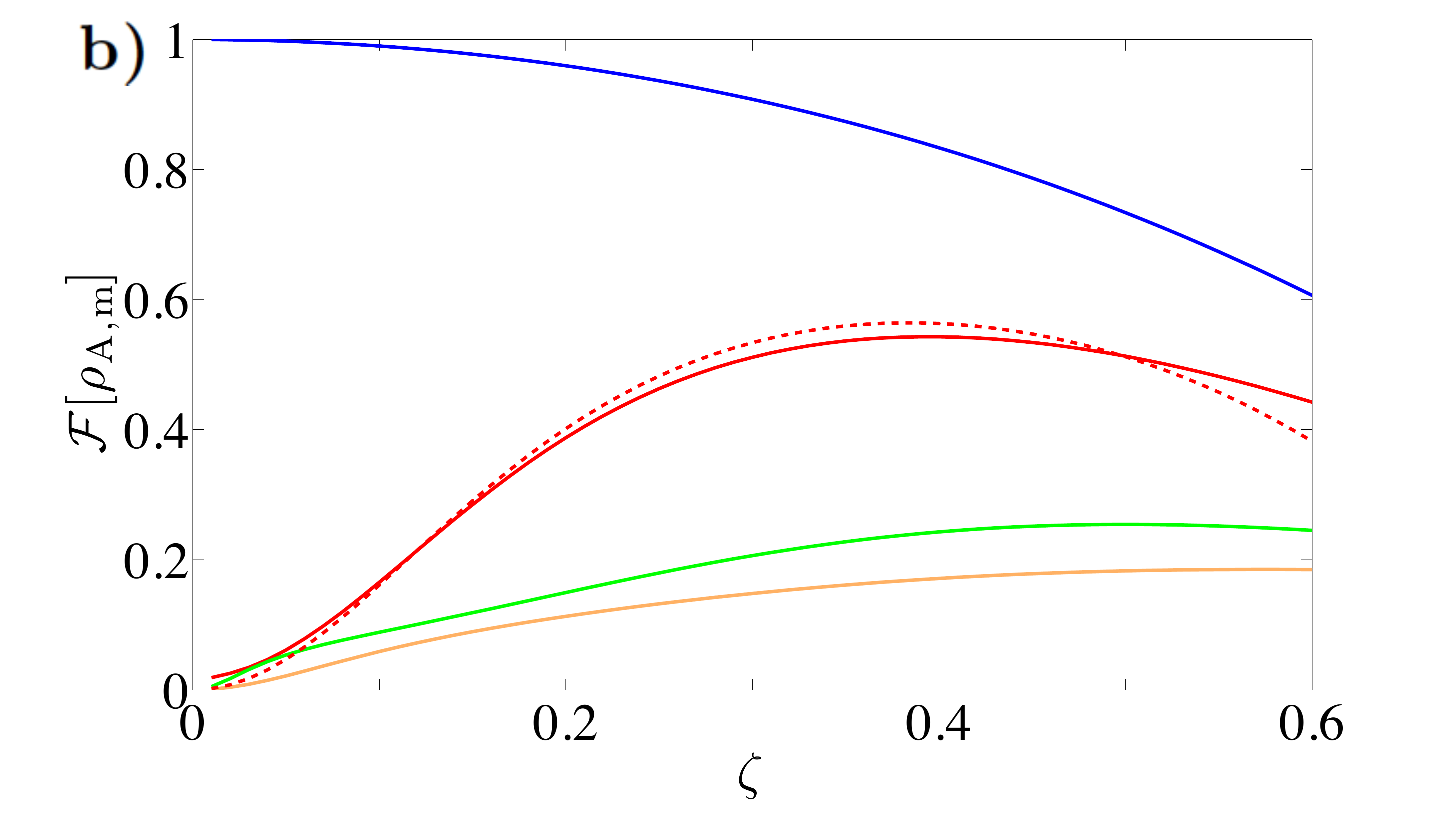}
\caption{(Color online) Fidelity for the preparation of a) $\tilde{\rho}_{\rm A}$ and b)  ${\rho}_{\rm A,m}$ as a function of $\zeta$ in the perturbative regime $\gamma_{\rm aux}/\gamma_{\rm eff}=0.1$. The different colors show various cooperativities and compare the numerical result (solid line) to the perturbative one (dashed line). \textit{Blue}: $\mathcal{C}_{\rm q}=\mathcal{C}_{\rm m}=\infty$, \textit{Red}: $\mathcal{C}_{\rm q}=\mathcal{C}_{\rm m}=100$, \textit{Green:} $\mathcal{C}_{\rm q}=\mathcal{C}_{\rm m}=20$, \textit{Orange:} $\mathcal{C}_{\rm q}=\mathcal{C}_{\rm m}=10$. } \label{Fig2}
\end{figure}
In this section we show how the previous results can be understood within perturbation theory. As the optimal case  $\gamma_{\rm eff}=\gamma_{\rm aux}$ cannot be described within perturbation theory, we focus on the perturbative limit $\gamma_{\rm aux}\ll\gamma_{\rm eff}$. We consider the gerneral jump operator $\hat{J}_1= \hat{\sigma}^++\eta\hat{\sigma}^-+\nu\bop-\zeta\bdop$ that prepares the qubit and the oscillator in the steady state given by eq.~\eqref{eq:rhossaux}. Perturbation theory shows that the maximal value for both $\mathcal{F}[\tilde{\rho}_{\rm A}]$ and $\mathcal{F}[\rho_{\rm A, m}]$  is obtained for $\nu=\eta=0$. We thus choose $\hat{J}_1=\hat{\sigma}^+-\zeta\bdop$ to compare with the numerical study~\footnote{Note that in principle the perturbation theory also applies for $\zeta\leq0$. However, care has to be taken as the range of validity of the analytical result depends on $\zeta$. This is because it is carried out assuming a finite-sized Hilbert space with maximal occupation number two for the harmonic oscillator. For $\zeta\leq 0$, high-occupation number states of the harmonic oscillator are excited more frequently than for $\zeta\geq0$.}. Within perturbation theory to first order, the steady state of $\mathcal{L}_0+\mathds{P}\left(\mathcal{L}_{\rm aux}+\mathcal{L}_{\rm q}+\mathcal{L}_{\rm m}\right)\mathds{P}$ is given by eq.~\eqref{eq:rhossaux}  with 
\be
\begin{split}
\alpha_{\rm aux}&=\frac{A\left(\gamma_{\rm m} \zeta^2+\gamma_{\rm aux}(1- \zeta^2 )^2\right)}{\gamma_{\rm q}A+ \gamma_{\rm m}B+\gamma_{\rm aux} C},\\
\beta_{\rm aux}&=\frac{ A(\gamma_{\rm q}+\gamma_{\rm m} \zeta^2 )+2 \gamma_{\rm aux} \zeta^4 (3- \zeta^2  (2-\zeta^2 ))}{\gamma_{\rm q}A+ \gamma_{\rm m}B+\gamma_{\rm aux} C},\label{eq:pop}
\end{split}
\ee
where $A=\left(3+4 \zeta ^2+\zeta^4\right)$, $B=2\zeta ^2 \left(3+4 \zeta^2+2 \zeta ^4\right)$, and $C=3-2 \zeta^2 +2\zeta^4-2\zeta^6+3\zeta^8$. 
This perturbative result is compared to a numerical simulation in Fig.~\ref{Fig2} for $\gamma_{\rm aux}\ll\gamma_{\rm eff}$. It is in good agreement with the numerical results, with an increasing deviation for lowered cooperativities. 

Also the results for the preparation of $\rho_{\rm SS,m}$ after carrying out the measurement as given by eq.~\eqref{eq:fock} can be understood within perturbation theory. $\alpha_{\rm m}$ and $\beta_{\rm m}$ are given by
\be
\begin{split}
\alpha_{\rm m}&=\frac{2\zeta^2\gamma_{\rm aux}(3-5\zeta^2+\zeta^4+\zeta^6)+2\zeta^4\gamma_{\rm m}(3+\zeta^2)}{A \gamma_{\rm q}+D\gamma_{\rm m}+E\gamma_{\rm aux}},\\
\beta_{\rm m}&=\frac{ A\gamma_{\rm q}+\gamma_{\rm m}\zeta^2(3+4\zeta^2+3\zeta^4)+2\gamma_{\rm aux}\zeta^4(3-2\zeta^2+\zeta^4)}{A \gamma_{\rm q}+D\gamma_{\rm m}+E\gamma_{\rm aux}},
\end{split}
\ee
with $D=\zeta^2(3+10\zeta^2+5\zeta^4)$ and $E=\zeta^2(6-4\zeta^2-2\zeta^4+4\zeta^6)$.
A numerical evaluation for different $\zeta$ as demonstrated in Fig.~\ref{Fig2}~b) shows that the perturbation theory is in accordance with the numerical prediction. 

\section{Coherent dynamics}\label{sec:coh}
Let us now consider the coherent dynamics corresponding to the regime given by $g_{\rm q}/|\Delta-\delta|\ll1,~g_{\rm m}/|\Delta-\omega|\ll1$. Eliminating the cavity mode from eq.~\eqref{eq:ham} (with $H_{\rm aux}=0$) gives
\be
H^{\rm eff}=\frac{\tilde\delta}{2}\hat{\sigma}_z+\tilde\omega\bdop\bop-g(\hat{\sigma}^+\bop+\hat{\sigma}^-\bdop)+\Omega(\hat{\sigma}^++\hat{\sigma}^-),\label{eq:heff}
\ee
where $\tilde{\delta}=\delta- 2 g_{\rm q}^2/(\Delta-\delta)$ and $\tilde{\omega}=\omega-2 g_{\rm m}^2/(\Delta-\omega)$ are the renormalized frequencies. The cavity-mediated coupling between the qubit and the mechanical oscillator is given by $g= g_{\rm q}g_{\rm m}(2\Delta-\omega-\delta)/\left[(\Delta-\delta)(\Delta-\omega)\right]$. 
In the good-cooperativity limit, several interesting phenomena can be observed.

First, the Hamiltonian of eq.~\eqref{eq:heff}, which is the well-known Jaynes-Cummings Hamiltonian, enables the preparation of arbitrary Fock states following the proposal of Law and Eberly~\cite{law96}. It relies on switching interaction strengths time-dependently by varying the laser intensities driving the different couplings. This requires $M$ steps for the preparation of arbitrary superposition states with maximal occupation number $M$. Therefore, all dissipation processes have to be slower than the coherent manipulation time, which is fulfilled  for $(g_{\rm q}g_{\rm m})/(\kappa\gamma_{\rm q(m)}),\mathcal{C}_{\rm q(m)} \gg M$.

Second, eq.~\eqref{eq:heff} predicts the occurence of blockade phenomena, a typical indicator of nonlinear behavior. Due to the presence of the qubit, the photon blockade~\cite{birnbaum07} is observable for $g_{\rm q} \gg\kappa,\gamma_{\rm q}$. Additionally, also the phonon blockade can be observed~\cite{rabl11, nunnenkamp11}: eliminating the atom to fourth order from eq.~\eqref{eq:heff} (justified for $g /|\delta-\omega|\ll1$ and $|\delta-\omega|>\gamma_{\rm q}$) yields an effective nonlinear Hamiltonian $H_{\rm phon}=\tilde{\omega}\bdop\bop+g^4/(\delta-\omega)^3(\bdop\bop)^2$. In addition, the good cooperativity  $g_{\rm q}g_{\rm m}/(\kappa\gamma_{\rm q(m)}),\mathcal{C}_{\rm q(m)} \gg 1$ ensures that the splittings effected by the nonlinear interaction are not smeared out by noise processes. 
\section{Many-body system}\label{sec:many}
An intriguing perspective in the field of optomechanics is to couple several nonlinear nanomechanical oscillators to realize a many-body system. This is required for quantum simulation~\cite{ludwig12} and might be particularly useful for the preparation of many-body states for quantum metrology. To achieve this goal, we propose to use a cavity to mediate the interaction between several mechanical oscillators and a single qubit. In order to realize $N$ nonlinear modes, we suggest to drive the mechanical frequencies stroboscopically. 
Any physical system with a tunable mechanical frequency, \eg, levitating dielectric spheres, can realize this protocol. In the following, the operators for each mechanical mode are termed $\bop_i$ ($i=1,...N$) with corresponding time-dependent frequencies $\omega_i(t)$ that are switched between a value on resonance $\omega_{\rm on}$ and off resonance $\omega_{\rm off}$. The case where the modulation of the couplings is achieved via a sinusoidal drive can be treated in full analogy.
 
The proposal requires the following conditions: (i) Due to the time-dependence of $\omega_i(t)$, also the operators $\bop_i (\bdop_i)$ are time-dependent. Requiring $\bop_i (\bdop_i)$ to be identical at the time of switching requires it to take place with a periodicity $\tau=2\pi n/\omega_{\rm off}$. (ii) The adiabatic elimination requires $g_{\rm m}/|\Delta-\omega_i(t)|, g_{\rm q}/|\Delta-\delta|\ll1$ (coherent dynamics) or $g_{\rm q (m)}/\kappa\ll1$ (dissipative dynamics). (iii) The stroboscopic switching has to be faster than the interaction between the different components of the system, therefore $g_{\rm m}\tau,~g_{\rm q}\tau\ll1$. (iv) The frequency change has to be the fastest time scale in the system, $(\omega_{\rm on}-\omega_{\rm off})\tau\gg1$. 
(v) The good cooperativity limit $\mathcal{C}_{\rm q}, \mathcal{C}_{\rm m}\gg1$ is necessary. 

\begin{figure}
\includegraphics[width=0.99\linewidth]{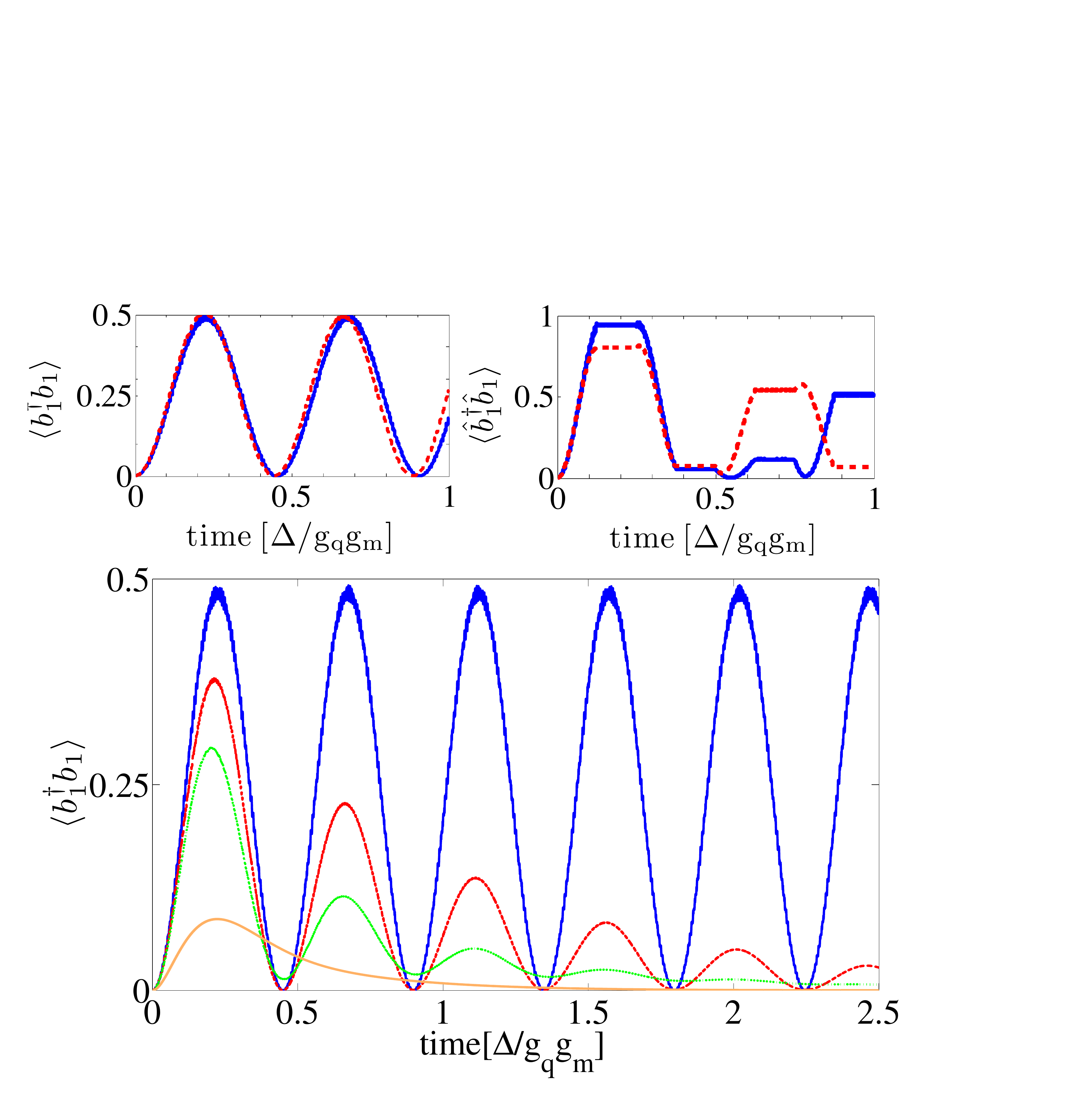}
\caption{(Color online) Dynamics of two stroboscopically-driven mechanical oscillators coupled to an initially excited qubit.
\textit{Top:} Comparison between the population of the mechanical mode under the full evolution (solid blue line) and the adiabatically eliminated one (dashed red line), all parameters are given in units of $g_{\rm q}g_{\rm m}/\Delta$ and no dissipation is included. \textit{Left:} Conditions (i)-(v) are fulfilled, \textit{Right:} Condition (i) is not fulfilled. \textit{Bottom}: Evolution under the influence of dissipation over time $[\Delta/g_{\rm m}g_{\rm q}]$ for different cooperativities: \textit{Solid Blue:} $\mathcal{C}_{\rm m}=\mathcal{C}_{\rm q}=\infty$, \textit{Dashed red:} $\mathcal{C}_{\rm m}=\mathcal{C}_{\rm q}=1000$, \textit{Dotted green:} $\mathcal{C}_{\rm m}=\mathcal{C}_{\rm q}=100$, \textit{Solid orange:} $\mathcal{C}_{\rm m}=\mathcal{C}_{\rm q}=10$. }\label{Fig4}
\end{figure}

In order to verify these conditions, we numerically simulate the stroboscopic driving of two oscillators as illustrated in Fig.~\ref{Fig4}. Initially, the qubit is in an excited state and it is shown that this excitation is coherently shifted to the mechanical oscillators and back to the qubit resulting in Rabi oscillations. We show in the upper panel that the stroboscopic driving is effective if conditions (i)-(v) are fulfilled. The robustness of the setup towards noise is illustrated in the lower panel, where the decay of the oscillations of the stroboscopically-driven system is analyzed for different cooperativities. It shows that the good-cooperativity limit is necessary, as otherwise oscillations decay rapidly. We plot the population of the first oscillator, as all other oscillators coupled to the qubit behave in full analogy. As shown below, the stroboscopic driving enables the individual addressability of each oscillator as opposed to the continous driving, where only the center-of-mass-mode is coupled.

\subsection{State preparation of the many-body system}\label{sec:many_prep}
Let us now translate the ideas for state preparation from the single-oscillator to the many-body case. To start the state-preparation in a well-defined state, each oscillator is cooled to its ground state via stroboscopic driving without coupling to the qubit. It can be shown that the effective coupling strength is $\propto g_{\rm m}/N$, and the light-scattering-induced dissipation scales $\propto \gamma_{\rm m}/N^2$,  rendering the cooperativity independent of the number of oscillators. Hence, the same conditions as in the single-particle case apply for ground-state cooling~\footnote{Note that in systems, where other sources of decoherence, \eg, heating through a direct thermal contact is dominant, the cooperativity might depend on the number of oscillators. This has to be taken into account accordingly.}. 
This can be used to prepare all the oscillators in their respective ground state,
\be
|\psi\rangle_{\rm ini}=\otimes_{i=1}^N|0\rangle_i \label{eq:ground_mb}.
\ee

One can now employ dissipative protocols to prepare interesting many-body non-Gaussian entangled states, \eg, the W-state
\be
|\psi\rangle_{\rm W}=\frac{1}{\sqrt{N}}(|10...0\rangle+...+|0...01\rangle).\label{eq:wstate}
\ee 
This can be achieved as follows. Starting from the ground state given by eq.~\eqref{eq:ground_mb}, all oscillators are tuned on resonance. In this case, the interaction between the qubit and the oscillators can be described in analogy to eqs.~\eqref{eq:leff} and \eqref{eq:laux} by the effective Liouvillians $\mathcal{L}_0^{\rm cm}[\rho]+\mathcal{L}_1^{\rm cm}[\rho]$ with jump operators $\hat{J}^{\rm cm}_0=\bop_{\rm cm}-\zeta\hat{\sigma}^-$ and $\hat{J}^{\rm cm}_1=\hat{\sigma}^+-\zeta\bdop_{\rm cm}$. Here $\bop_{\rm cm}=\sum_{i=1}^N\bop_i/\sqrt{N}$ denotes the center-of-mass operator. In full analogy to the single-oscillator case, the system can be dissipatively prepared in a Fock state of the center-of-mass-motion of the mechanical oscillators by performing a measurement of the qubit's state followed by postselection. This leads to the W-state given by eq.~\eqref{eq:wstate}, namely $|\psi\rangle_{\rm W} = \hat b^{\dagger}_\text{cm} \ket{\psi}_\text{ini}$. The fidelity to prepare the system in this dark state is thus given by the fidelity of the protocol for single oscillators and can reach, \eg, $\mathcal{F}[\ket{\psi}_{\text{W}} \bra{\psi}]\approx0.83$ for $\gamma_{\rm aux}=\gamma_{\rm eff}$ and $\mathcal{C}_{\rm q}=\mathcal{C}_{\rm m}=100$.

\subsection{Coherent state preparation of $N$ mechanical oscillators}\label{sec:law_many}
In the following we develop a method for coherent state preparation of a system consisting of $N$ mechanical oscillators and a single qubit. Our approach is based on a protocol proposed by Law and Eberly, see~\cite{law96}, that has already been discussed in Sec.~\ref{sec:coh}. Here, we provide its extension to $N$-body systems. The goal is to determine the full time evolution $U(t_{\rm fin})$ that prepares a system, initially in its ground state $|\psi\rangle_{\rm ini}$ (eq.~\eqref{eq:ground_mb}), in a target state $|\psi\rangle_{\rm target}$. The key tool of~\cite{law96} is to realize that this evolution operator may be obtained by solving the equations of motion of the inverse evolution $U(-t_{\rm fin})$ given by
\be
|g\rangle\otimes|\psi\rangle_{\rm ini}=U(-t_{\rm fin})|g\rangle\otimes|\psi(t_{\rm fin})\rangle_{\rm target}.
\ee
It transfers the system from the target state $|\psi(t_{\rm fin})\rangle_{\rm target}$ to its ground state. 

In the many-body case the goal is to evolve the initial state, eq.~\eqref{eq:ground_mb}, to the general Fock state
\be
|\psi(t_{\rm fin})\rangle_{\rm target}=\sum_{n_1=0,...,n_N=0}^M c_{n_1...n_N}|n_1,...,n_N\rangle,\label{eq:target}
\ee
with maximal occupation number $M$ for each of the $N$ oscillators at time $t_{\rm fin}$. In order to extend the Law-Eberly approach, it is essential to address each of the states separately. This requires a Hamiltonian that is only on resonance with one specific state at a time.

As it has been demonstrated previously, the time-dependent switching of the frequencies of the mechanical oscillators enables single-oscillator addressability. The Hamiltonian of the system is given by the many-body extension of eq.~\eqref{eq:heff} with $H_{\rm aux}=0$. Being off-resonant, the other oscillators may be adiabatically eliminated during the manipulation of the $j$th oscillator, which gives
\be
\begin{split}
H^{\rm eff}_{j}=&\frac{\delta}{2}\hat{\sigma}_z+\sum_{i\neq j}^N l_i n_i \hat{\sigma}_z+\omega_j\bdop_j\bop_j\\
&-g_j(t)(\hat{\sigma}^+\bop_j+\hat{\sigma}^-\bop_j)+\Omega_j(t)(\hat{\sigma}^++\hat{\sigma}^-),
\end{split}\label{eq:eberly_mb}
\ee
with $l_i=-2g_i^2/(\delta-\omega_i)$. The second term in eq.~\eqref{eq:eberly_mb} describes the renormalization of the atomic frequency determined by the occupation number $n_i$ of all off-resonant oscillators. It has to be taken into account when turning the $j$th oscillator on resonance with the atom. This additional renormalization shift enables a unique addressing of each state of the many-body system provided that $\sum_{i\neq j} l_i(n_i-n'_i)=0$ iff $n_i=n_i',  \forall i$.

Hence, the operation that prepares the $j$th oscillator in the desired state is given by
\be
U_j=U_j^{(n_1=M,...,n_N=M)}...U_j^{(n_1=1,...,n_N=1)},\label{eq:uni_j}
\ee
where the dots in the multiplication account for all possible permutations of occupation numbers of the off-resonant oscillators. Each $U_j^{(n_1,...,n_N)}$ performs the Law-Eberly protocol on the $j$th oscillator under the condition that the other oscillators are in state $|n_1,...,n_{\rm N}\rangle$. The mechanism is subsequently applied to all oscillators yielding the full time evolution $U(t_{\rm fin})=U_N...U_{\rm 1}$. In general, the maximal number of necessary steps for the preparation of an arbitrary state eq.~\eqref{eq:target} is given by
\be
\#(\rm{steps})=\sum_{i=0}^{N-1}(M+1)M.
\ee
It increases from $M$ steps for the preparation of the $M$th Fock state in the single-oscillator case to at most $M^N$ steps in the many-body case.

As an illustration, let us consider the necessary steps for the preparation of
\be
|\psi\rangle_{\rm spec}=\frac{1}{\sqrt{3}}\left(|0,5,0\rangle+|1,5,10\rangle+|1,1,1\rangle\right).\label{eq:example}
\ee
We consider the inverse evolution $U(-t_{\rm fin})=U_3^{\dagger}U_2^{\dagger}U_1^{\dagger}$ preparing eq.~\eqref{eq:example} in the ground state. Applying $U_1^{\dagger}=U_1^{\dagger,(n_2=1,n_3=1)}U_1^{\dagger,(n_2=5,n_3=10)}$, as defined in eq.~\eqref{eq:uni_j}, requires 2 steps and prepares the first oscillator in the ground state. The subsequent preparation of the second oscillator is performed by $U_2^{\dagger}=U_1^{\dagger,(n_1=0,n_3=0)}U_1^{\dagger,(n_1=0,n_3=1)}U_1^{\dagger,(n_1=0,n_3=10)}$ and requires 11 steps. Finally, we apply $U_3^{\dagger}=U_3^{\dagger,(n_1=0,n_2=0)}$ to the third oscillator, which requires 10 steps. In total, the preparation of eq.~\eqref{eq:example} can be achieved in 23 steps and the specific operators may be determined in full analogy to the single-particle case~\cite{law96}.

\section{Conclusion and Outlook}\label{sec:conc}

In this article we investigated the nonlinear physics obtained by adding a single qubit to a cavity-nanomechanical system. The cavity mode mediates the coupling between the qubit and the mechanical oscillator. In optomechanics this can be achieved by placing a single two-level atom inside the same cavity.  We have shown how to prepare non-Gaussian states of the mechanical oscillator, such as superpositions of Fock states. We focused on dissipative methods where the fast decay of the cavity mode is exploited for state preparation. It has been demonstrated that engineering the environment strongly increases the fidelity to prepare non-Gaussian states to values close to one. In order to gain a better understanding of the results, we developed a perturbation theory for degenerate Liouvillians and derived an analytic description of the occurring phenomena. Additionally, we proposed a method to prepare many-body Hamiltonians with N nonlinearities by adding N mechanical oscillators to the cavity containing a single qubit. This was achieved by stroboscopically driving the mechanical oscillators. Albeit the focus of this article is on state-preparation, the N-body system realized via the stroboscopic driving enables the simulation of several interesting many-body problems~\cite{porras04}. Since the cavity mediates the couplings between all oscillators, one could use this setup to implement a Bose-Hubbard model with long interaction range. Other applications include the study of superradiance and dissipative phase transitions.

Funding by the Elite Network of Bavaria (ENB) project QCCC (A.C.P.), the EU project SIQS and Fundaci\'o Catalunya-La Pedrera is gratefully acknowledged.

\end{document}